\documentclass[conference]{IEEEtran}

\usepackage[english]{babel}
\usepackage[utf8x]{inputenc}
\usepackage{amsmath}
\usepackage{graphicx}
\usepackage{xspace}
\usepackage{hyperref}
\usepackage{varioref}
\usepackage{balance}
\usepackage{subfig}
\usepackage{fixltx2e}

\usepackage{xcolor}
\definecolor{dark-red}{rgb}{0.4,0.15,0.15}
\definecolor{dark-blue}{rgb}{0.15,0.15,0.4}
\definecolor{medium-blue}{rgb}{0,0,0.5}
\hypersetup{
    colorlinks, linkcolor={dark-red},
    citecolor={dark-blue}, urlcolor={medium-blue}
}

\newcommand{\ie}{\textit{i.e.}\xspace}
\newcommand{\eg}{\textit{e.g.}\xspace}
\newcommand{\etc}{\textit{etc.}\xspace}
\newcommand{\etal}{\textit{et al.}\xspace}
\newcommand{\adhoc}{\textit{ad hoc}\xspace}
\newcommand{\kil}{\texttt{kill}\xspace}
\newcommand{\wai}{\texttt{wait}\xspace}
\newcommand{\sus}{\texttt{suspend}\xspace}
\newcommand{\res}{\texttt{resume}\xspace}

\newcommand{\plotwidth}{.8\columnwidth}

\title{OS-Assisted Task Preemption for Hadoop}
\author{\IEEEauthorblockN{Mario Pastorelli, Matteo Dell'Amico, Pietro
    Michiardi}\IEEEauthorblockA{EURECOM, France}}

\begin{document}
\maketitle

\begin{abstract}
This work introduces a new task preemption primitive for Hadoop, that allows tasks to be suspended and resumed exploiting existing memory management mechanisms readily available in modern operating systems. Our technique fills the gap that exists between the two extremes cases of killing tasks (which waste work) or waiting for their completion (which introduces latency): experimental results indicate superior performance and very small overheads when compared to existing alternatives.

\end{abstract}

\section{Introduction}
\label{sec:intro}

Data-intensive scalable computing (DISC) frameworks, such as Hadoop \cite{hadoop} and Spark~\cite{spark}, have received great attention by both industry and academia, as they allow to design and execute scalable algorithms to process large amounts of data, with the ultimate goal of better understanding users, business processes, and services, in a variety of application domains.

In many situations, organizations resort to \emph{separate clusters} to make sure that data exploration and algorithm tuning jobs do not interfere with well-tested, production ones. Indeed, it is common to differentiate jobs that are essential for the business of an organization that runs them, from other jobs that can be seen as ``best-effort'': the latter category should not use resources that high-priority jobs would need.

Clearly, physical partitioning of jobs and clusters entails high overheads in terms of system administration, and an inefficient use of cluster resources. A more flexible and efficient solution is to consolidate clusters, and (manually) define job priorities to inform resource allocation. Unfortunately, while systems such as Hadoop provide ways of setting priorities for jobs, such techniques are imperfect since the only mechanisms available to implement them impose a choice between waiting for low-priority tasks\footnote{In Hadoop, and in DISC systems in general, a \emph{task} is a unit of processing work which is performed on a single machine. A typical Hadoop task can last tens of seconds or minutes.} to finish before resources can be granted to high-priority ones, or killing such tasks, thereby wasting the work they have done so far. 

The endeavor of this work is to provide a new, \emph{transparent} solution, that allows preempting running tasks, in order to \emph{run high-priority tasks with low latencies} while \emph{avoiding wasting work}. As we discuss in Section~\ref{sec:related}, priorities that are manually set by developers are not the only use case 
that benefits from an efficient preemption primitive: preemption is important for size-based and deadline-based scheduling, and for enforcing fairness in resource allocation. Still in Section~\ref{sec:related}, we discuss preemption primitives that are available in Hadoop, with their merits and their shortcomings.

Our solution, elucidated in Section~\ref{sec:method}, uses operating system (OS) mechanisms to suspend and resume tasks, which -- in current DISC frameworks -- are standard UNIX processes. Coherently with the implementation of Hadoop, which uses standard POSIX signals to communicate with processes, we perform suspension and resuming with, respectively, the \texttt{SIGTSTP} and \texttt{SIGCONT} signals.

With our approach, the state of tasks is implicitly saved by the operating system, and kept in memory. If not enough physical memory is available for running tasks at any moment, the OS paging mechanisms saves the memory allocated to the suspended tasks in the swap area. This step avoids overheads due to systematic serialization and deserialization, and is generally rare in systems with abundant memory.\footnote{Ananthanarayanan \etal report that ``the median and $95^{th}$ percentile memory utilizations [in Facebook clusters] are 19\% and 42\%, respectively.''~\cite{ananthanarayanan2012pacman}.}

As the experimental results of Section~\ref{sec:experiments} show, our preemption primitive outperforms current approaches in both our performance goals: providing low latencies to high-priority tasks and avoiding redundant work. Even when the available memory is limited, the overhead due to paging is very small.

We further note that our preemption primitive has implications on both implementing Hadoop schedulers and writing MapReduce programs. Hadoop schedulers have a better way to perform task preemption, but they should decide \emph{which} tasks to evict; those who are writing MapReduce programs should consider optimizing them in order to minimize the amount of allocated memory. These issues are discussed in Section~\ref{sec:discussion}.

In Section~\ref{sec:conclusion}, we conclude and discuss further research.

\section{Related Work}
\label{sec:related}

Preemption is an important concept in scheduling in general, and in
addition to the manual priority settings we described in the
Introduction, there are several use cases in a system such as Hadoop
that can benefit from such a primitive. Job schedulers, like the Hadoop
FAIR and Capacity schedulers, can use preemption to warrant
fairness~\cite{zaharia2009job}: if a job starves due to long-running
tasks of another job, these latter may be preempted. In deadline
scheduling~\cite{kc2010scheduling}, preemption can be used to make
sure that jobs that are close to the deadline are run as soon as
possible. Size-based schedulers~\cite{wolf2010flex,pastorelli2013hfsp}
in general attribute priorities to jobs according to a virtual or real
size, and preemption can guarantee that higher-priority jobs are
allowed to run earlier.

Currently, two preemption strategies are available for Hadoop. One technique is to wait for tasks that should be preempted to complete: this is done using the \wai{} strategy. Another approach is to kill tasks, using the \kil{} primitive. Clearly, the first policy has the shortcoming of introducing large latencies for high-priority tasks, while the second one wastes work done by killed tasks. We refer to the work by Cheng \etal \cite{cheng2011mitigating} for an approach that strives to mitigate the impact of the \kil{} strategy by adopting an appropriate eviction policy (\ie, choosing which tasks to kill). In Section~\ref{sec:experiments}, we compare our new preemption primitive to \wai{} and \kil{}.

A recent preemption mechanism for Hadoop is Natjam~\cite{cho2013natjam}: unlike in 
our work, where we use the OS to perform process suspension and resuming, Natjam operates at the ``application layer'', and saves counters about task progress, which allow to resume tasks by fast-forwarding to their previous states. Since the state of the Java Virtual Machine (JVM) is lost, however, Natjam cannot be applied seamlessly to arbitrary tasks: indeed, many MapReduce programming patterns involve keeping track of a state within the task JVM~\cite{lin2010data}; this problem is exacerbated by the fact that many MapReduce jobs are created by high-level languages
such as Apache Pig~\cite{olston2008pig} or Apache Hive~\cite{thusoo2009hive}: jobs compiled by these frameworks are highly likely to make use of these ``tricks'', which hinders the application of Natjam. 

Natjam proposes to handle such stateful tasks with hooks that systematically serialize and deserialize task state. Besides requiring manual intervention to support suspension, this approach has the drawback of always requiring the overhead for serialization, writing to disk, and deserialization of a state that could be large. In contrast, our approach does not incur in a systematic serialization overhead, since it relies on OS paging to swap to disk the state of the tasks, \emph{if} and \emph{when} needed. 

\section{OS-assisted Task Preemption}
\label{sec:method}

We now describe our preemption primitive, that implements task suspension and resume operations. First, we outline how process suspension and memory paging work in modern operating systems.  
Then, we present the implementation of our preemption mechanism. Note that this work focuses solely on preemption primitives, and glosses over \textit{task eviction policies} that are within the scope of a job and task scheduler.

\subsection{Suspension and Paging in the OS}
\label{sec:suspension}

Here we provide a synthetic description of the way OSes perform memory management, which motivate our design and implementation. A more in-depth description of such mechanisms can be found, for example, in the work of Arpaci-Dusseau~\cite[Chapters 20 and 21]{ostep}.

In general, system RAM is occupied by file-system (disk) cache and runtime memory allocated by processes (including map/reduce tasks); when RAM is full -- for whatever reason -- the OS needs to \emph{evict} pages from memory, either by reclaiming space (and evict pages) from the file-system cache or by \emph{paging out} runtime memory to the swap area. Since Hadoop workloads involve large sequential reads from disks, it is a best practice to configure the Linux kernel to give precedence to runtime memory, always evicting file-system cache first~\cite{cdh4-tips}. The system therefore only pages out runtime memory to avoid ``out of memory'' conditions, \ie when the memory allocated by \textit{running} processes exceeds the physical RAM. 

To decide which pages to swap to disk, OSes generally employ a policy which is a variant of least-recently-used (LRU)~\cite{split-lru}; \emph{clean} pages -- \ie, those that have not been modified since the last time they have been read from disk -- do not need to be written and get prioritized when performing eviction. Page-out operations are generally clustered to improve disk throughput (and amortize on seek
costs) by writing multiple pages to disk in a batch. These implementation policies ensure that paging is efficient and with small overheads, especially if a suspended processes leads to swapping. Most importantly for our case, pages from suspended processes are evicted before those from running ones.

We recall that it is necessary to make sure that the aggregate memory size for all processes -- both running and suspended -- does not exceed the size of the swap space on disk, because in such a case the operating system would be forced to kill processes. Since Hadoop tasks can only allocate a limited amount of memory, this can be ensured by configuring the scheduler so that also the number of suspended tasks
per task-tracker is limited.

\noindent \textbf{Thrashing.} Paging, in general, is not problematic unless \emph{thrashing} happens, a phenomenon where data is continuously read from and written to swap space~\cite{denning1968thrashing} on disk. Thrashing is caused by a \emph{working set} -- \ie, the set of pages accessed by running programs -- which is larger than main memory.

In Hadoop, thrashing is avoided because two mechanisms are in place: \textit{i)} the number of running tasks per machine is limited (and controlled via a configuration parameter); and \textit{ii)} the heap space size that a given task can allocate is limited (and also controlled via configuration). Proper Hadoop configuration can thus
limit working set size and avoid thrashing.

The aforementioned mechanisms prevent thrashing in the same way even when suspension is used. Memory allocated by suspended processes is \emph{outside the working set} and hence \emph{cannot cause thrashing}; pages allocated for the suspended processes are paged out and in \emph{at most once}, respectively after suspension and resuming. Thrashing could only happen if a given job is continuously suspended and resumed by the scheduling mechanism: the moderate cost of a suspend-resume cycle can be thus multiplied by the number of cycles. A reasonable scheduler implementation should take into account that suspending and resuming a job has a cost, and should take measures to avoid paying it too often.

\subsection{Implementation Details}
\label{sec:implementation}

The concepts that we illustrate here are valid for both Hadoop 1 \cite{hadoop}, which is the most widely used Hadoop implementation in production, Hadoop 2, which uses a new infrastructure for resource negotiation called YARN~\cite{vavilapalli2013yarn}, and even other frameworks such as Spark~\cite{spark}. Currently, our implementation targets Hadoop version 1.

Our preemption primitive exposes an API that can be used both by users on the
command line and by schedulers.  Mirroring the implementation of the \kil{} primitive in Hadoop, we introduce \textit{i)} new messages between the JobTracker (a centralized machine responsible for keeping track of system state and scheduling) and TaskTrackers (machines responsible for running Map/Reduce tasks), and \textit{ii)} new identifiers for task states in the JobTracker. 

\noindent \textbf{JobTracker.} Hadoop has a ``heartbeat'' mechanism where, at fixed intervals and every time a task finishes, TaskTrackers inform the JobTracker about their state.

As soon as the JobTracker receives the command to suspend a task from
the user or the scheduler, that task is marked as being in a
\texttt{MUST\_SUSPEND} state. At the following heartbeat from the
involved TaskTracker, the JobTracker piggybacks the command to
suspend the task. The following heartbeat notifies the JobTracker
whether the task has been suspended -- which triggers entering the
\texttt{SUSPENDED} state in the JobTracker -- or whether it completed
in the meanwhile.

Analogous steps are taken to resume tasks, exchanging
appropriate messages and handling the \texttt{MUST\_RESUME} state,
returning the state to \texttt{RUNNING} when the process is over.

\noindent \textbf{TaskTracker.} In Hadoop, Map and Reduce tasks are regular Unix processes running in child JVMs spawned by the TaskTracker. This means that they can safely be handled with the POSIX signaling infrastructure. In particular, to suspend and resume tasks, our preemption primitive
uses the standard POSIX \texttt{SIGTSTP} and \texttt{SIGCONT} signals.

These signals are used because (unlike \texttt{SIGSTOP}) they allow handlers to be written to manage external state, \eg, when closing and reopening network connections.

\noindent \textbf{Job and Task Scheduler.} 
We factor out the role of task eviction policies implemented by the
scheduler, which are not the focus of this work, by building a new
scheduling component for Hadoop -- a dummy scheduler -- which dictates
task eviction according to static configuration files. This allows to
specify, using a series of simple triggers, which jobs/tasks are run
in the cluster and which are preempted. In addition to executing jobs
and preempting tasks with our \sus{}/\res{} primitives, the dummy
scheduler also allows using the \kil{} primitive and to \wai, for the
purpose of a comparative analysis.

\section{Experimental Evaluation}
\label{sec:experiments}

In our experiments, we evaluate preemption primitives in terms of the latency they introduce and the amount of redundant work they require. We show that our approach outperforms other preemption primitives and has a small overhead both when jobs are lightweight in terms of memory, and when they are memory-hungry.

\begin{figure}
\includegraphics[width=\columnwidth]{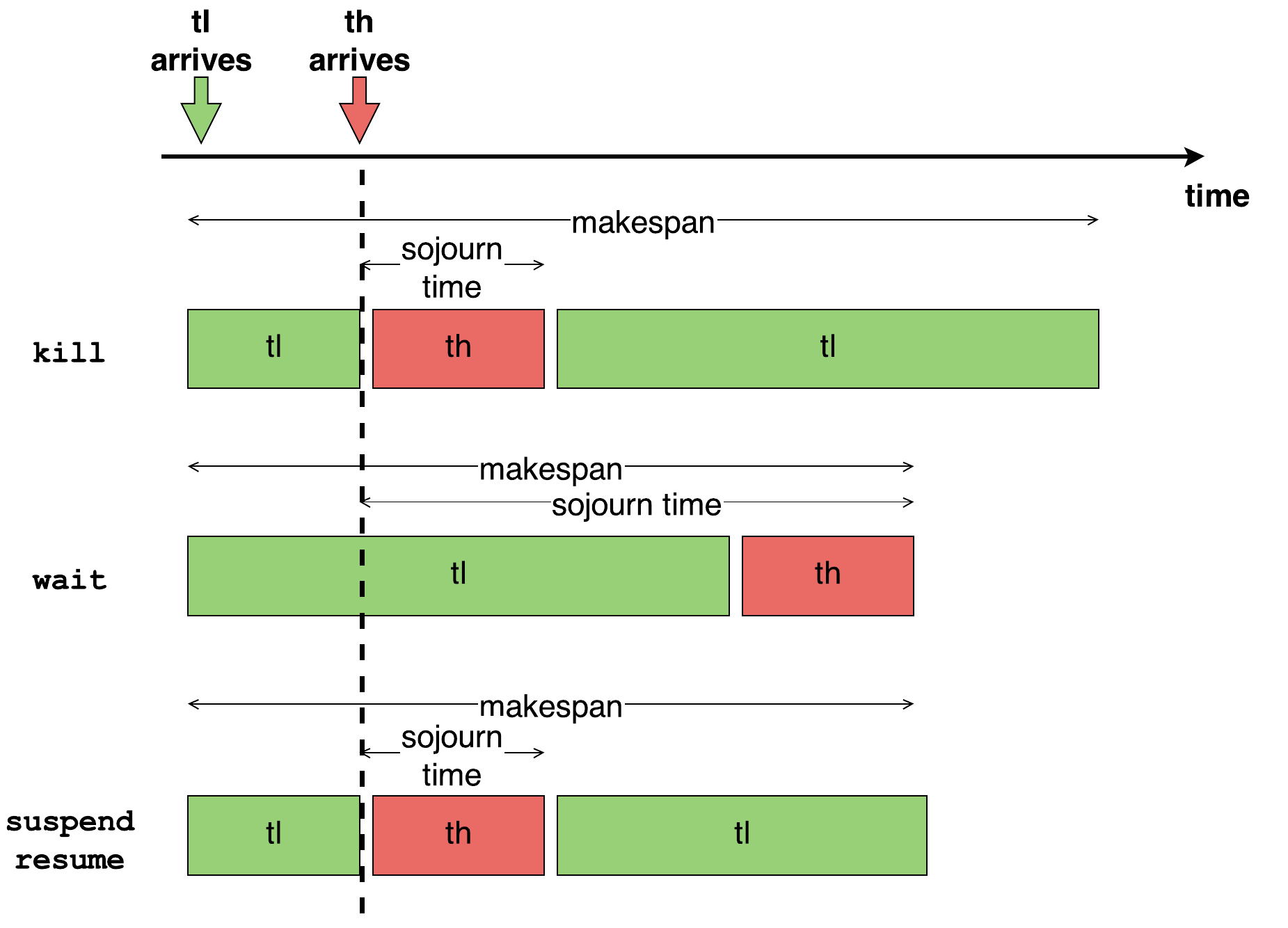}
\caption{Task execution schedules.}
\label{fig:task-diagram}
\end{figure}

\subsection{Experimental Setup}
\label{sec:settings}

Our \sus{}/\res{} primitives operate at the task level, and behave in the same way for both Map and Reduce tasks. We evaluate the behavior of the system in a simple setup: our dummy scheduler runs two single-task, map-only jobs, called $t_h$ and $t_l$ ($h$ and $l$ stand for high and low priority respectively). $t_l$ processes a single-block file stored on HDFS, with size 512 MB; $t_h$ processes single HDFS input block of size 512 MB. Both jobs run synthetic mappers, which read and parse the randomly generated input.
We remark that this setup is analogous to the one used by Cho \etal, who evaluated their preemption primitive using similar synthetic jobs created by the SWIM workload generator~\cite{chen2011case}.

In our experiments, our dummy scheduler preempts the low-priority task $t_l$ after it has reached a completion rate $r\%$ (\ie, $r\%$ of the input tuples have been processed) and grants the task slot to the high priority task $t_h$. Once $t_h$ is completed, the scheduler resumes $t_l$, which can complete as well.

Next, we evaluate the behavior of our \sus{}/\res{} preemption
mechanism against the two baseline primitives available in Hadoop:
\wai{} and \kil{}. When waiting, task $t_h$ is simply executed after
$t_l$ completes; when killing, task $t_l$ is killed as soon as $t_h$
is scheduled, and $t_l$ is rescheduled from scratch after the
completion of $t_h$. This simple experimental setup is illustrated in
Figure~\vref{fig:task-diagram}.

According to Hadoop configuration best practices, in our experimental setup we prioritize runtime memory over disk cache and therefore limit
swapping, as discussed in Section~\ref{sec:suspension}, by setting the Linux \texttt{swappiness} parameter to 0.

\subsection{Performance Metrics}

Our goals are ensuring low latency for high-priority tasks, and avoid wasting work: we quantify them, respectively, with the \emph{sojourn time of $t_h$} and the \emph{makespan} of the workload. \emph{Sojourn Time of $t_h$} is the time that elapses between the moment $t_h$ is submitted and when it completes; \emph{makespan} is the time that passes between the moment in which the first task $t_l$ is submitted and when \emph{both} tasks are complete.

\subsection{Results}

We focus on experimental results in case of light-weight tasks. This is the standard case for ``functional'', stateless, mappers and reducers.
In this case, the amount of memory that tasks allocate is essentially due to the Hadoop execution engine (\ie, JVM, I/O buffers, overhead due to sorting, \etc). 

Stateful mappers and reducers, instead, can allocate non-negligible amounts of memory; we thus complement our experiments by studying our performance metrics and overheads for memory-hungry jobs, which represent a worst-case scenario for our preemption primitive.

All our results are obtained by averaging 20 experiment runs; we omit error bars for readability: in all data points reported, minimum and maximum values measured are within 5\% of the average values.

\begin{figure*}
 \centering \begin{tabular}{cc}
    \begin{minipage}[t]{0.49\textwidth}
      \begin{center}
        \subfloat[Sojourn time of $t_h$]{
          \label{fig:th-sojourn}
          \includegraphics[width=\plotwidth]{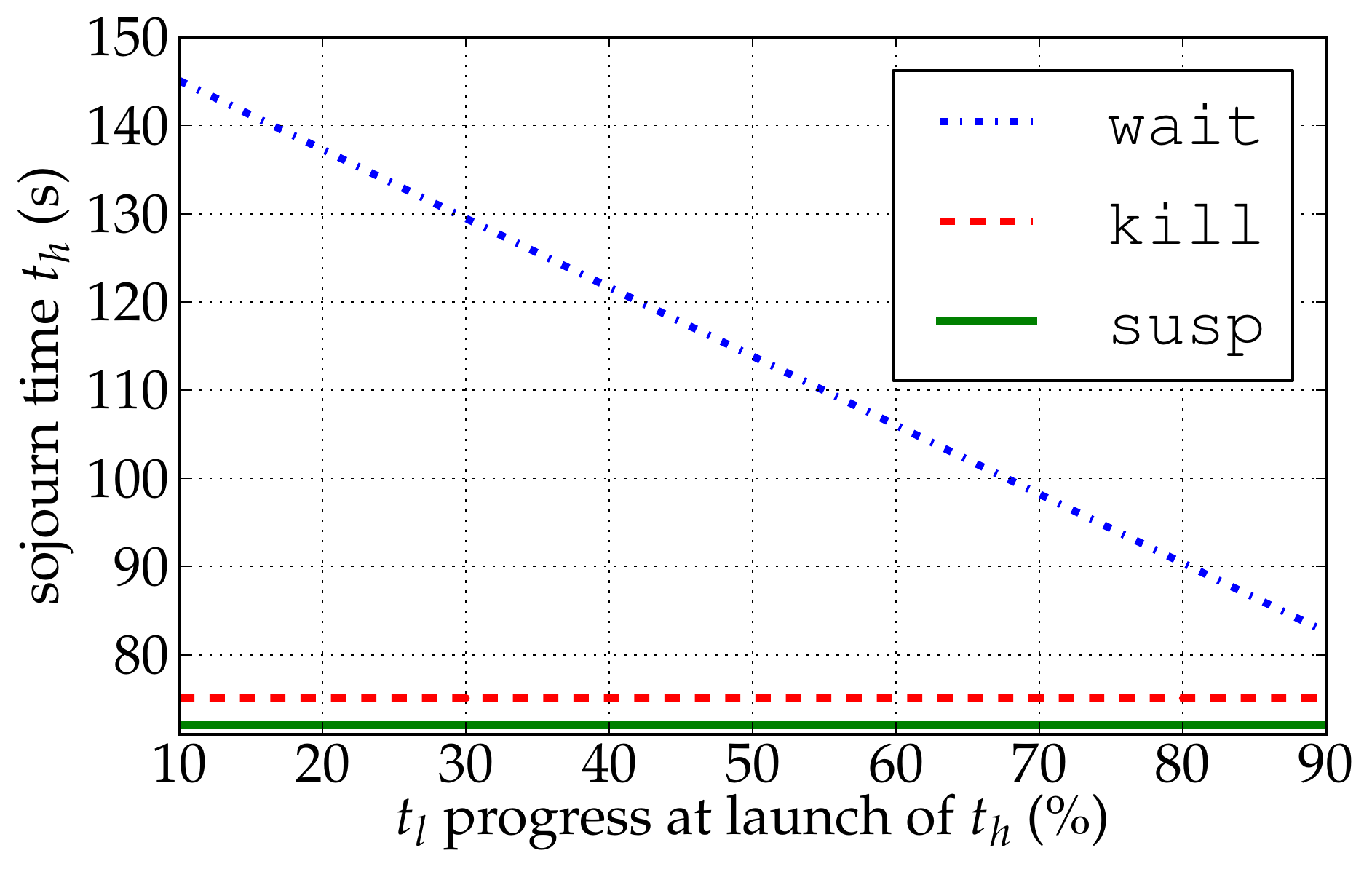}
        }
      \end{center}
    \end{minipage}

    &
    
    \begin{minipage}[t]{0.49\textwidth}
      \begin{center}
        \subfloat[Makespan]{
          \label{fig:makespan}
          \includegraphics[width=\plotwidth]{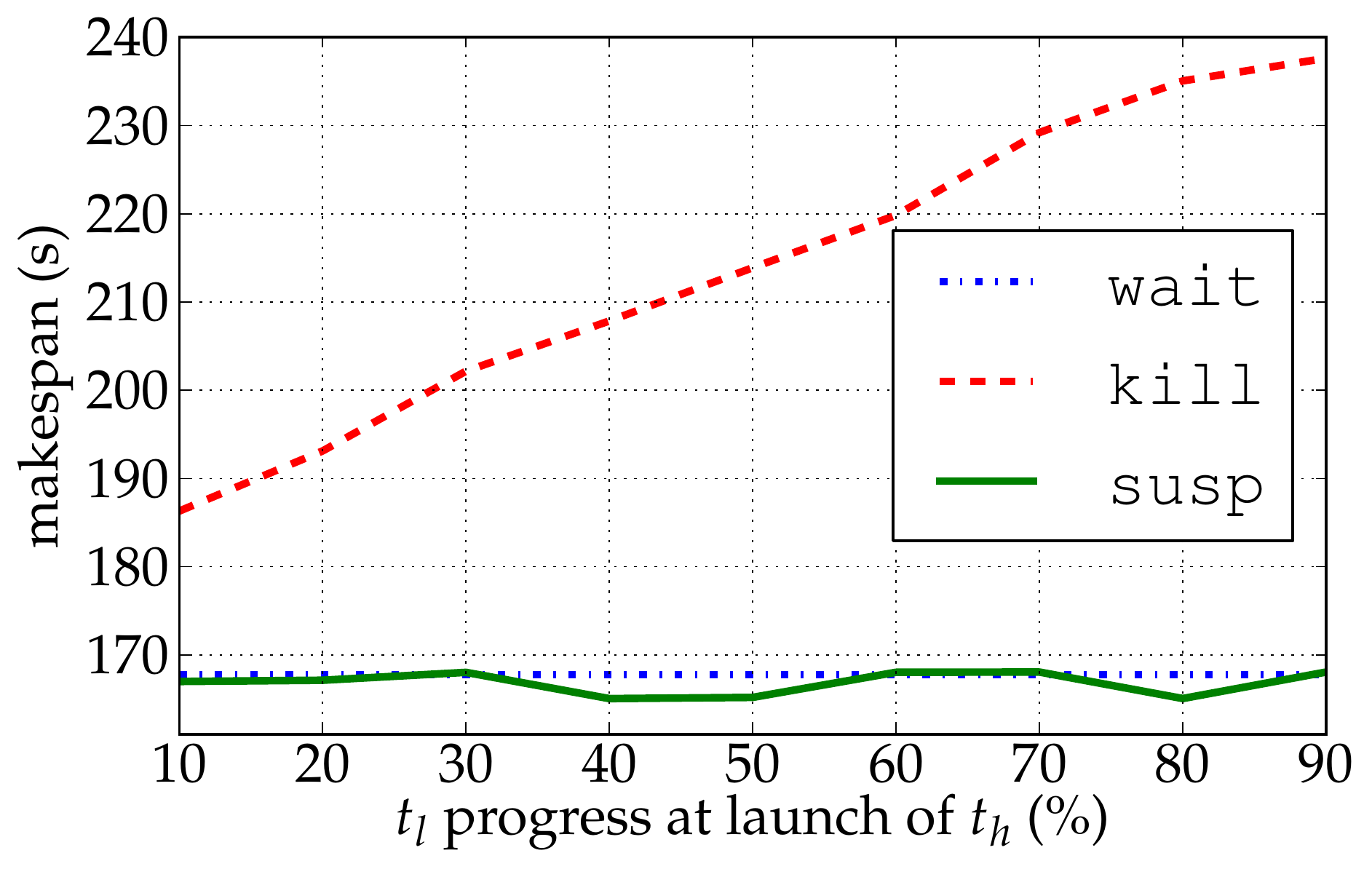}
        }
      \end{center}
    \end{minipage}

  \end{tabular}

  \caption{Baseline experiments: a comparison of the three preemption primitives with light-weight tasks.}
\end{figure*}

\noindent \textbf{Baseline Experiments.} Figure~\vref{fig:th-sojourn} illustrates the sojourn time of $t_h$: the arrival rate of $_h$ is a parameter defined as a function of $t_l$ progress, as shown on the x-axis. 

The \kil{} and our \sus{}/\res{} primitives achieve small sojourn times, as opposed to \wai{}, in particular when $t_h$ arrives early. However, they both incur in some overheads: \kil{} runs a cleanup task to remove temporary outputs of the killed task; \sus{}/\res{} may slow down $t_h$ in case paging out memory occupied by $t_l$ is needed. In our baseline setup, both jobs are light-weight, hence the suspended process resides only in memory. This explains the small advantage for our mechanism, which outperforms all other primitives even when $t_h$ arrives at 90\% completion rate of task $t_l$.

Figure~\vref{fig:makespan} illustrates our results for the makespan metric, using the same setup described above. In this case, the makespan is heavily affected by a preemption primitive that wastes work. The \wai{} policy, at the cost of delaying $t_h$, avoids supplementary work and achieves a small makespan; the \kil{} primitive, instead, wastes all the work done by $t_l$ before preemption. Finally, our preemption primitive behaves similarly to the \wai{} policy, despite the possible overhead due to page-out/page-in cycles.

For light-weight jobs, we conclude that our primitive is superior to both alternatives, as both sojourn times and makespan are small. We note that the authors of Natjam measured an overhead of around
7\% in terms of makespan, in similar experimental settings as ours. Our findings suggest that the overhead in our case is negligible.

\noindent \textbf{Worst-Case Experiments.} The experiments discussed above are valid for simple implementations of Map and Reduce tasks, that carry out stateless computations on their input. Stateful tasks can, however, allocate memory, which may force the OS to swap. Since clusters often have plentiful available memory~\cite{ananthanarayanan2012pacman}, such a situation is unlikely to be frequent. 
However, we still consider a ``worst case'' scenario to stress our primitive: both $t_l$ and $t_h$ allocate a large amount of memory (2 GB in our case; we note that this requires an \adhoc change to the Hadoop configuration since Hadoop jobs are not generally allowed to allocate such an amount of memory). This value makes sure that, when running a single task the system does not have to recur to swap;\footnote{The physical memory of our system is 4 GB; the rest of the memory is needed by the Hadoop framework and by the operating system services.} conversely, when the two tasks are present in the system at the same time, one of them is forced to page out memory. We ensure that tasks allocate memory and that the OS marks pages as ``dirty'', by writing random values to all memory at task startup, and reading them back when finalizing the tasks.

\begin{figure*} \centering \begin{tabular}{cc}
    \begin{minipage}[t]{0.49\textwidth}
      \begin{center}
        \subfloat[Sojourn time of $t_h$]{
          \label{fig:th-sojourn-himem}
          \includegraphics[width=\plotwidth]{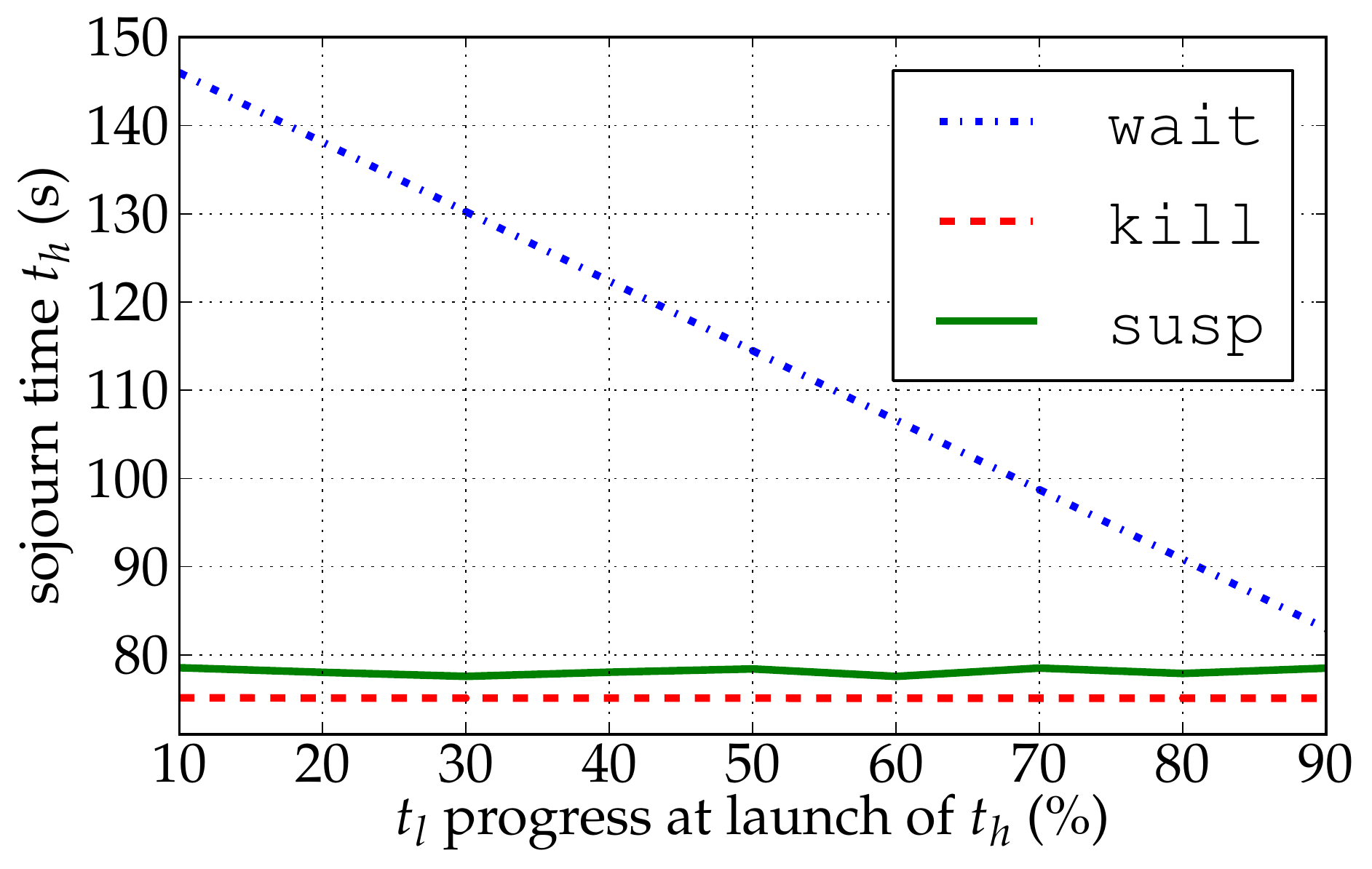}
        }
      \end{center}
    \end{minipage}

    &
    
    \begin{minipage}[t]{0.49\textwidth}
      \begin{center}
        \subfloat[Makespan]{
          \label{fig:makespan-himem}
          \includegraphics[width=\plotwidth]{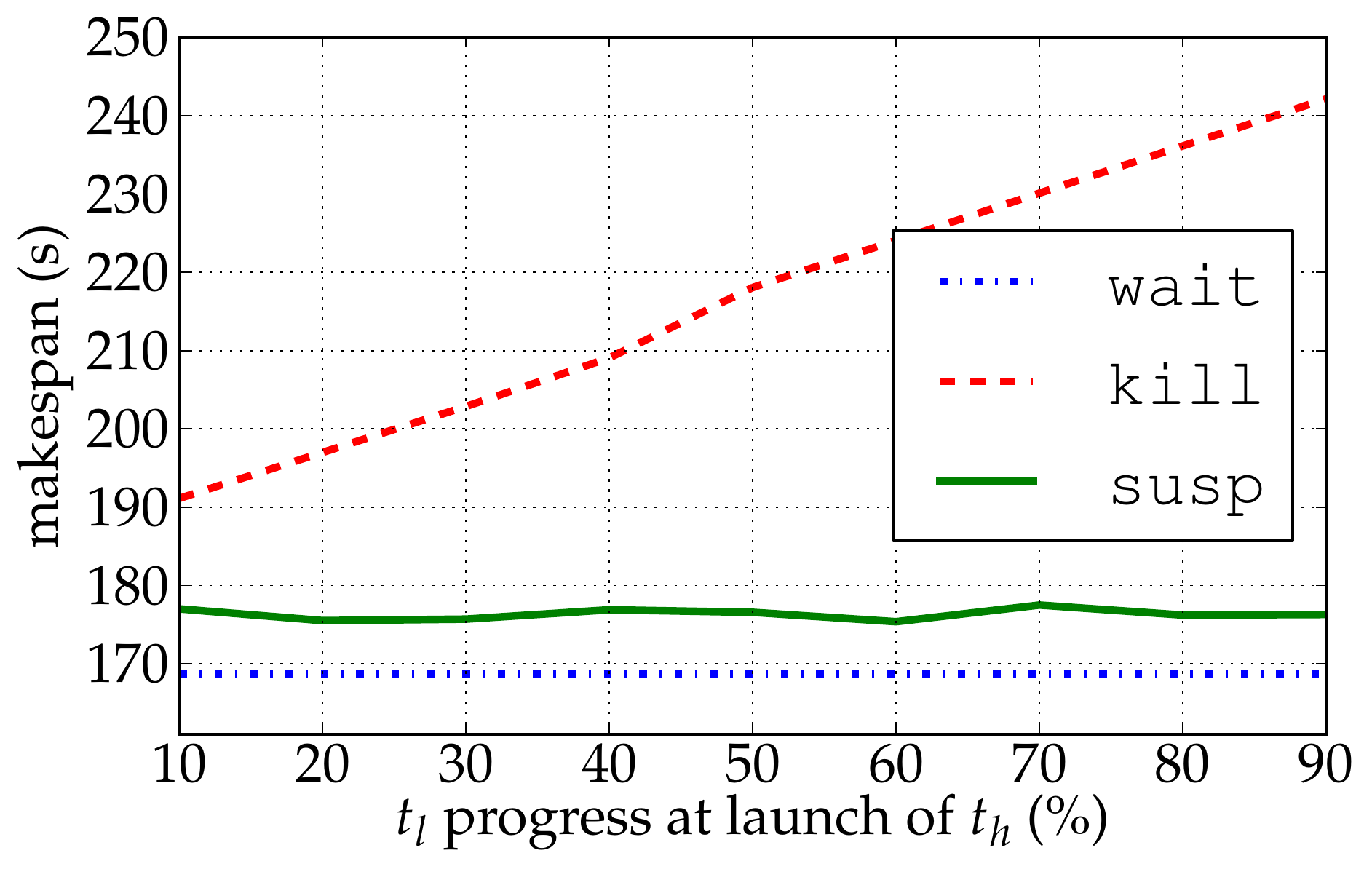}
        }
      \end{center}
    \end{minipage}

  \end{tabular}

  \caption{Worst-case experiments: a comparison of the three preemption primitives with memory-hungry tasks.}
\end{figure*}

Figures~\ref{fig:th-sojourn-himem} and~\vref{fig:makespan-himem} present the sojourn time and the makespan for the worst-case experimental setup. While our preemption primitive still outperforms both alternatives with respect to both metrics, it is possible to notice that the overheads related to paging are visible: with respect to the sojourn time, the \kil{} primitive achieves a slightly lower value; similarly, the \wai{} primitive achieves slightly smaller makespan. Overall, the overhead due to our preemption primitive is marginal: we further investigate and quantify it in the next section.

\noindent \textbf{Impact of Memory Footprint.} We now focus on a detailed analysis of the overheads imposed by the OS paging mechanism on the performance of our preemption primitive. To do so, we vary the amount of memory a task allocates in the setup phase.\footnote{This is where, generally, auxiliary data structures are created to maintain an internal state in a task.} 
In our experiments $t_l$ allocates 2.5 GB of memory, and we parametrize over the amount of memory $t_h$ allocates. For each experimental run, we measure the number of bytes swapped by the process executing $t_l$, and compute the degradation of sojourn time and makespan compared to the \kil{} and \wai{} primitives, respectively.

\begin{figure}
\centering
\includegraphics[width=\plotwidth]{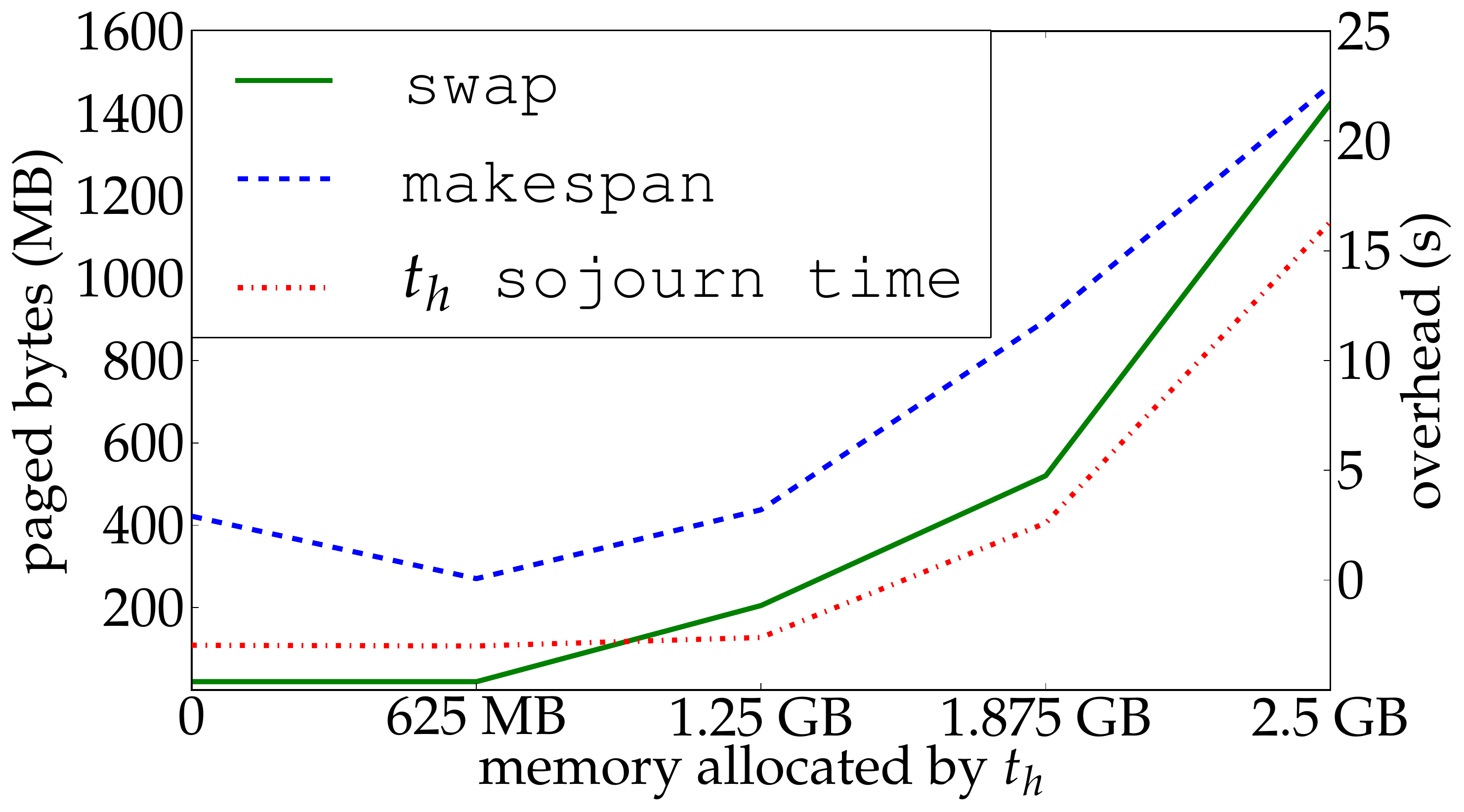}
\caption{Overheads when varying memory usage.}
\label{fig:memory}
\end{figure}

Figure~\ref{fig:memory} indicates that the overheads due to paging are roughly linearly correlated to the amount of data swapped to disk. For the sojourn time, our preemption primitive degrades when $t_h$ allocates more than 1.5 GB of RAM: in the worst-case, sojourn time is 20\% larger than with the \kil{} primitive. Similarly, for the makespan, our mechanism degrades when $t_h$ allocates more than 1.3 GB: in the worst-case, makespan is 12\% larger than with the \wai{} primitive. Finally, we note that swapped data grows more than linearly because of an approximate implementation of the page replacement algorithm in Linux (and other modern OSes), which can lead to more swapping than strictly necessary~\cite[Chapter 17]{Bovet:2005:ULK:1077084}.

\section{Discussion}
\label{sec:discussion}

We now elaborate more on the implications of the new preemption primitive we introduce in this work.

\subsection{Scheduling and Eviction Strategies}
\label{sec:eviction}

As we discussed in the Introduction, our \sus{} primitive gives one
more opportunity to the developers of schedulers, in order to perform
more efficient preemption. As we have shown with the results of
Section~\ref{sec:experiments}, our primitive generally performs close
to optimally in most cases; however, for freshly started tasks, it may be
preferable to use the \kil{} primitive, and for tasks that are very
close to completion it may be better to simply \wai{} for them to
finish.

\noindent \textbf{Task Eviction Policies.} An important topic that falls under the responsibility of the schedulers is to decide \emph{which} task(s) to evict once a
high-priority job needs time to execute. Cho
\etal~\cite{cho2013natjam} propose to suspend tasks that are closest
to completion, in order to have all tasks of a job as close to each
other as possible, resulting in a good influx on job sojourn times. If
the goal is instead to avoid redundant work and reduce makespan,
another possible strategy may aim to suspend tasks with smaller memory
footprints, which reduces overheads according to our experimental results.

\noindent \textbf{Resume Locality.} \label{sec:reduce-locality} In our implementation, a suspended process can only be resumed on the
same machine it was suspended on. If the same task gets scheduled on a
different machine, it has to be restarted from scratch, losing work
done so far: in that case, the \sus{} is effectively analogous to a
delayed \kil{}. We call this issue \emph{resume locality} due to its
similarity with the \emph{data locality} issue -- \ie, the problem of
running mappers on the machines that have local copies of data.

Hadoop schedulers generally handle data locality by using the simple
technique of delay scheduling~\cite{zaharia2010delay}: waiting a fixed
amount of time before scheduling non-local copies of data. Only if
that threshold is exceeded, a non-local mapper is run. The same
technique can be used for our resume locality issue.

Analogously to our approach, Natjam only supports local resumes. As a
future improvement, the authors suggest moving the
checkpoints used to mark task state and reduce inputs over the
network; a similar approach could be taken also in our case, using
process migration facilities such as CRIU~\cite{criu}. However,
extreme care should be taken before attempting to use such a non-local
resume in particular for reducers, since the cost of moving non-local
inputs can be very large.

\subsection{Implications on Task Implementation}

In most cases, our \sus{}/\res{} mechanism is transparent towards the
implementation, and task implementations that correctly handle error
conditions and the possibility of being killed by the scheduler will
also handle suspension correctly. However, we add a few notes
regarding tasks with external state and ways in which task
implementation can control the memory footprint.

\noindent \textbf{External State.} In some cases, Hadoop jobs can interact with the external world through more than inputs and outputs: they can use network connections
and/or use ``Hadoop Streaming'', whereby arbitrary executables can be
used as mappers or reducers, interacting with the Hadoop framework
through Unix pipes. In these cases, there are interactions that happen
outside the control of Hadoop; in the most common case, external
software would correctly pause waiting for the next input from a
suspended task; however, when the interaction happens with a complex
program, the fact that they correctly handle suspended programs should
be tested.

\noindent \textbf{Controlling Memory Footprint.} We have seen that the memory footprint allocated by a process has an impact on the overheads due to suspension; when writing task implementations, it is good measure to take this into account and
optimize for lower memory footprints.

Java garbage collectors differ in the way they are implemented: some
of them release memory to the OS when they stop using it, others do
not~\cite{java-gc-comparison}. It is therefore a good idea to
configure Java to use a garbage collector that does release memory,
such as the new G1 implementation~\cite{g1}. It is also possible to
hint the garbage collector to run using \texttt{System.gc()}; this is
advisable after disposing of large objects in memory.

\balance

\section{Conclusion}
\label{sec:conclusion}
In this work we presented a new task preemption primitive that improves over existing techniques to perform both manual and automatic scheduling of Hadoop jobs. 

The gist of our preemption primitive was to make use of the memory management mechanisms readily available in the OS to perform task suspension and resuming. Motivated by the limitations of current approaches to task suspension -- that implement preemption at the ``application level'' -- we argued that an OS-assisted approach could provide a general preemption mechanism that seamlessly supported a variety of workloads, including stateful tasks.

We implemented our preemption primitive for Hadoop, and discussed how to modify its core components to take into acccount the suspended state of a task, and the signalling mechanisms to trigger task suspension and resume.

Finally, we implemented a simple Hadoop scheduler that allowed us to focus on the goals of our comparative analysis of preemption mechanisms. In our experiments, we glossed over the details and variety of task eviction policies implemented by standard schedulers, and we compared the performance of the \kil{}, \wai{} and \sus{}/\res{} mechanisms, paying particular attention in quantifying the overheads due to the OS memory management mechanisms. We did so in a variety of experimental settings, including worst-case scenarios of memory-hungry Hadoop jobs.
We showed that our technique fills the gap that exists between the two extremes cases of killing tasks (which waste work) or waiting for their completion (which introduces latency): performance is near-optimal, while overhead is small in most cases.

We have preliminary results showing that our preemption primitive
performs well in the context of HFSP, our size-based scheduler for
Hadoop~\cite{DBLP:journals/corr/abs-1302-2749}. Our next steps involve
a comprehensive study of task eviction policies implemented in
standard Hadoop schedulers that make use of our preemption primitive,
a thorough experimental campaign with realistic workloads, and the
application of our technique to additional DISC frameworks, such as
Apache Spark.

\bibliographystyle{IEEEtran}
\bibliography{references}

\begin{thebibliography}{10}
\providecommand{\url}[1]{#1}
\csname url@samestyle\endcsname
\providecommand{\newblock}{\relax}
\providecommand{\bibinfo}[2]{#2}
\providecommand{\BIBentrySTDinterwordspacing}{\spaceskip=0pt\relax}
\providecommand{\BIBentryALTinterwordstretchfactor}{4}
\providecommand{\BIBentryALTinterwordspacing}{\spaceskip=\fontdimen2\font plus
\BIBentryALTinterwordstretchfactor\fontdimen3\font minus
  \fontdimen4\font\relax}
\providecommand{\BIBforeignlanguage}[2]{{%
\expandafter\ifx\csname l@#1\endcsname\relax
\typeout{** WARNING: IEEEtran.bst: No hyphenation pattern has been}%
\typeout{** loaded for the language `#1'. Using the pattern for}%
\typeout{** the default language instead.}%
\else
\language=\csname l@#1\endcsname
\fi
#2}}
\providecommand{\BIBdecl}{\relax}
\BIBdecl

\bibitem{hadoop}
Apache, ``Hadoop,'' \url{http://hadoop.apache.org/}.

\bibitem{spark}
------, ``Spark,'' \url{http://spark.incubator.apache.org/}.

\bibitem{ananthanarayanan2012pacman}
G.~Ananthanarayanan, A.~Ghodsi, A.~Wang, D.~Borthakur, S.~Kandula, S.~Shenker,
  and I.~Stoica, ``Pacman: Coordinated memory caching for parallel jobs,'' in
  \emph{USENIX NSDI}, 2012.

\bibitem{zaharia2009job}
\BIBentryALTinterwordspacing
M.~Zaharia, ``Job scheduling with the fair and capacity schedulers,'' 2009.
  [Online]. Available:
  \url{https://trac.nchc.org.tw/grid/raw-attachment/wiki/jazz/09-09-22/FairScheduler_MateiZaharia_Cloudera.pdf}
\BIBentrySTDinterwordspacing

\bibitem{kc2010scheduling}
K.~Kc and K.~Anyanwu, ``Scheduling {H}adoop jobs to meet deadlines,'' in
  \emph{CloudCom}.\hskip 1em plus 0.5em minus 0.4em\relax IEEE, 2010.

\bibitem{wolf2010flex}
J.~Wolf, D.~Rajan, K.~Hildrum, R.~Khandekar, V.~Kumar, S.~Parekh, K.-L. Wu, and
  A.~Balmin, ``Flex: A slot allocation scheduling optimizer for mapreduce
  workloads,'' in \emph{Middleware 2010}.\hskip 1em plus 0.5em minus
  0.4em\relax Springer, 2010.

\bibitem{pastorelli2013hfsp}
M.~Pastorelli, A.~Barbuzzi, D.~Carra, M.~Dell'Amico, and P.~Michiardi,
  ``{HFSP}: size-based scheduling for {H}adoop,'' in \emph{Big Data}.\hskip 1em
  plus 0.5em minus 0.4em\relax IEEE, 2013.

\bibitem{cheng2011mitigating}
L.~Cheng, Q.~Zhang, and R.~Boutaba, ``Mitigating the negative impact of
  preemption on heterogeneous mapreduce workloads,'' in \emph{Proceedings of
  the 7th International Conference on Network and Services Management}.\hskip
  1em plus 0.5em minus 0.4em\relax International Federation for Information
  Processing, 2011, pp. 189--197.

\bibitem{cho2013natjam}
B.~Cho, M.~Rahman, T.~Chajed, I.~Gupta, C.~Abad, N.~Roberts, and P.~Lin,
  ``Natjam: Design and evaluation of eviction policies for supporting
  priorities and deadlines in mapreduce clusters,'' in \emph{SoCC}.\hskip 1em
  plus 0.5em minus 0.4em\relax ACM, 2013.

\bibitem{lin2010data}
J.~Lin and C.~Dyer, ``Data-intensive text processing with {MapReduce},''
  \emph{Synthesis Lectures on Human Language Technologies}, 2010.

\bibitem{olston2008pig}
C.~Olston, B.~Reed, U.~Srivastava, R.~Kumar, and A.~Tomkins, ``Pig latin: a
  not-so-foreign language for data processing,'' in \emph{SIGMOD}.\hskip 1em
  plus 0.5em minus 0.4em\relax ACM, 2008.

\bibitem{thusoo2009hive}
A.~Thusoo, J.~S. Sarma, N.~Jain, Z.~Shao, P.~Chakka, S.~Anthony, H.~Liu,
  P.~Wyckoff, and R.~Murthy, ``Hive: a warehousing solution over a map-reduce
  framework,'' \emph{PVLDB}, vol.~2, no.~2, 2009.

\bibitem{ostep}
\BIBentryALTinterwordspacing
R.~H. Arpaci-Dusseau and A.~C. Arpaci-Dusseau, \emph{Operating Systems: Three
  Easy Pieces}, 2013. [Online]. Available:
  \url{http://pages.cs.wisc.edu/~remzi/OSTEP/}
\BIBentrySTDinterwordspacing

\bibitem{cdh4-tips}
\BIBentryALTinterwordspacing
``{CDH} 4 installation guide -- tips and guidelines,'' Cloudera. [Online].
  Available:
  \url{http://www.cloudera.com/content/cloudera-content/cloudera-docs//CDH4/4.2.0/CDH4-Installation-Guide/cdh4ig\_topic\_11\_6.html}
\BIBentrySTDinterwordspacing

\bibitem{split-lru}
\BIBentryALTinterwordspacing
``Page replacement design,'' LinuxMM. [Online]. Available:
  \url{http://linux-mm.org/PageReplacementDesign}
\BIBentrySTDinterwordspacing

\bibitem{denning1968thrashing}
P.~J. Denning, ``Thrashing: Its causes and prevention,'' in \emph{Fall Joint
  Computer Conference}.\hskip 1em plus 0.5em minus 0.4em\relax ACM, 1968.

\bibitem{vavilapalli2013yarn}
V.~K. Vavilapalli, A.~C. Murthy, C.~Douglas, S.~Agarwal, M.~Konar, R.~Evans,
  T.~Graves, J.~Lowe, H.~Shah, S.~Seth \emph{et~al.}, ``Apache {H}adoop {Y}arn:
  Yet another resource negotiator,'' in \emph{SoCC}.\hskip 1em plus 0.5em minus
  0.4em\relax ACM, 2013.

\bibitem{chen2011case}
Y.~Chen, A.~Ganapathi, R.~Griffith, and R.~Katz, ``The case for evaluating
  mapreduce performance using workload suites,'' in \emph{MASCOTS}.\hskip 1em
  plus 0.5em minus 0.4em\relax IEEE, 2011.

\bibitem{Bovet:2005:ULK:1077084}
D.~Bovet and M.~Cesati, \emph{Understanding The Linux Kernel}.\hskip 1em plus
  0.5em minus 0.4em\relax O'Reilly \& Associates Inc, 2005.

\bibitem{zaharia2010delay}
M.~Zaharia, D.~Borthakur, J.~Sen~Sarma, K.~Elmeleegy, S.~Shenker, and
  I.~Stoica, ``Delay scheduling: a simple technique for achieving locality and
  fairness in cluster scheduling,'' in \emph{EuroSys}.\hskip 1em plus 0.5em
  minus 0.4em\relax ACM, 2010.

\bibitem{criu}
\BIBentryALTinterwordspacing
``{CRIU}: Checkpoint/restore in userspace.'' [Online]. Available:
  \url{http://criu.org}
\BIBentrySTDinterwordspacing

\bibitem{java-gc-comparison}
\BIBentryALTinterwordspacing
S.~Krause, ``{JDK} 7 {GC} behavior: To free or not to free,'' August 2011.
  [Online]. Available: \url{http://www.stefankrause.net/wp/?p=14}
\BIBentrySTDinterwordspacing

\bibitem{g1}
\BIBentryALTinterwordspacing
M.~Beckwith, ``{G1}: One garbage collector to rule them all,'' July 2013.
  [Online]. Available:
  \url{http://www.infoq.com/articles/G1-One-Garbage-Collector-To-Rule-Them-All}
\BIBentrySTDinterwordspacing

\bibitem{DBLP:journals/corr/abs-1302-2749}
M.~Pastorelli, A.~Barbuzzi, D.~Carra, M.~Dell'Amico, and P.~Michiardi,
  ``Practical size-based scheduling for {MapReduce} workloads,'' \emph{CoRR},
  vol. abs/1302.2749, 2013.

\end{thebibliography}

\end{document}